\documentclass[aps,prl,superscriptaddress,reprint,showpacs,longbibliography,amsmath,amssymb]{revtex4-1}
\usepackage[utf8]{inputenc}
\usepackage[T1]{fontenc}
\usepackage{microtype}
\usepackage{graphicx}
\usepackage{mathbbol}
\usepackage{amssymb}
\usepackage{comment}
\usepackage[usenames,dvipsnames,svgnames,table]{xcolor}
\usepackage[colorlinks,linkcolor=blue,citecolor=blue,urlcolor=blue]{hyperref}

\def\equationautorefname~#1\null{Eq.~(#1)\null}

\newcommand{\braket}[1]{\ensuremath{\langle{#1}\rangle}}
\newcommand{\bra}[1]{\langle #1 |}
\newcommand{\ket}[1]{| #1 \rangle}

\begin{document}
	
	\title{Tensor network simulation of non-Markovian dynamics in organic polaritons}
	
	\author{Javier del Pino}
	\affiliation{Departamento de Física Teórica de la Materia Condensada and Condensed Matter Physics Center (IFIMAC), Universidad Autónoma de Madrid, E-28049 Madrid, Spain}
	\author{Florian A. Y. N. Schröder}
	\affiliation{Cavendish Laboratory, University of Cambridge, J. J. Thomson Avenue, Cambridge, CB3 0HE, UK}
	\author{Alex W. Chin}
	\affiliation{Institut des NanoSciences de Paris, Sorbonne Université, 4 place Jussieu, boîte courrier 840, 75252, PARIS Cedex 05}
	\affiliation{Cavendish Laboratory, University of Cambridge, J. J. Thomson Avenue, Cambridge, CB3 0HE, UK}
	\author{Johannes Feist}
	\email{johannes.feist@uam.es}
	\affiliation{Departamento de Física Teórica de la Materia Condensada and Condensed Matter Physics Center (IFIMAC), Universidad Autónoma de Madrid, E-28049 Madrid, Spain}
	\author{Francisco J. Garcia-Vidal}
	\email{fj.garcia@uam.es}
	\affiliation{Departamento de Física Teórica de la Materia Condensada and Condensed Matter Physics Center (IFIMAC), Universidad Autónoma de Madrid, E-28049 Madrid, Spain}
	\affiliation{Donostia International Physics Center (DIPC), E-20018 Donostia/San Sebastián, Spain}
	
	\begin{abstract}
		We calculate the exact many-body time dynamics of polaritonic states supported
		by an optical cavity filled with organic molecules.  Optical, vibrational and
		radiative processes are treated on an equal footing employing the Time-Dependent
		Variational Matrix Product States algorithm. We demonstrate signatures of
		non-Markovian vibronic dynamics and its fingerprints in the far-field photon
		emission spectrum at arbitrary light-matter interaction scales, ranging from the
		weak to the strong coupling regimes. We analyze both the single and
		many-molecule cases, showing the crucial role played by the collective motion of
		molecular nuclei and dark states in determining the polariton dynamics and the
		subsequent photon emission.
	\end{abstract}
	
	\maketitle
	
	Organic polaritons, formed upon hybridization of optical electromagnetic (EM)
	modes and Frenkel excitons of organic molecules~\cite{Lidzey1998, Bellessa2004,
		Dintinger2005,Torma2015,Sanvitto2016}, exist as a threefold mixture of photonic, electronic
	and phononic excitations. As a result, modified nuclear effects in polaritons
	leads to tailored material~\cite{Orgiu2015,Feist2015,Schachenmayer2015} and
	chemical~\cite{Hutchison2012,Galego2015,Herrera2016,Bennett2016,Flick2018Abinitio}
	properties. In this regard, experiments have underlined the impact of vibronic states
	in the polariton time evolution, with examples in thermalization and cooling of
	room $T$ condensates~\cite{Rodriguez2013,Plumhof2014,Daskalakis2014},
	thresholds in organic lasers~\cite{Kena-Cohen2010,Ramezani2017Plasmon} and
	lifetimes~\cite{Schwartz2013,Vasa2013,Eizner2017}. First studies on the impact
	of vibrations on organic polariton dynamics relied on the Fermi's golden rule
	within a Markovian approach~\cite{Litinskaya2004,Michetti2009}, which
	neglects correlations and interplay between electronic and vibrational
	excitations. More recently, the influence of vibronic states in the
	spectroscopic properties of these systems~\cite{Herrera2017Theory,Zeb2018} has been
	analyzed using the Holstein-Tavis-Cummings (HTC) model, in which the complex spectral
	density of the vibronic modes in organic molecules is modelled by a single
	phonon mode.
	
	
	\begin{figure}
		\includegraphics[width=0.9\linewidth]{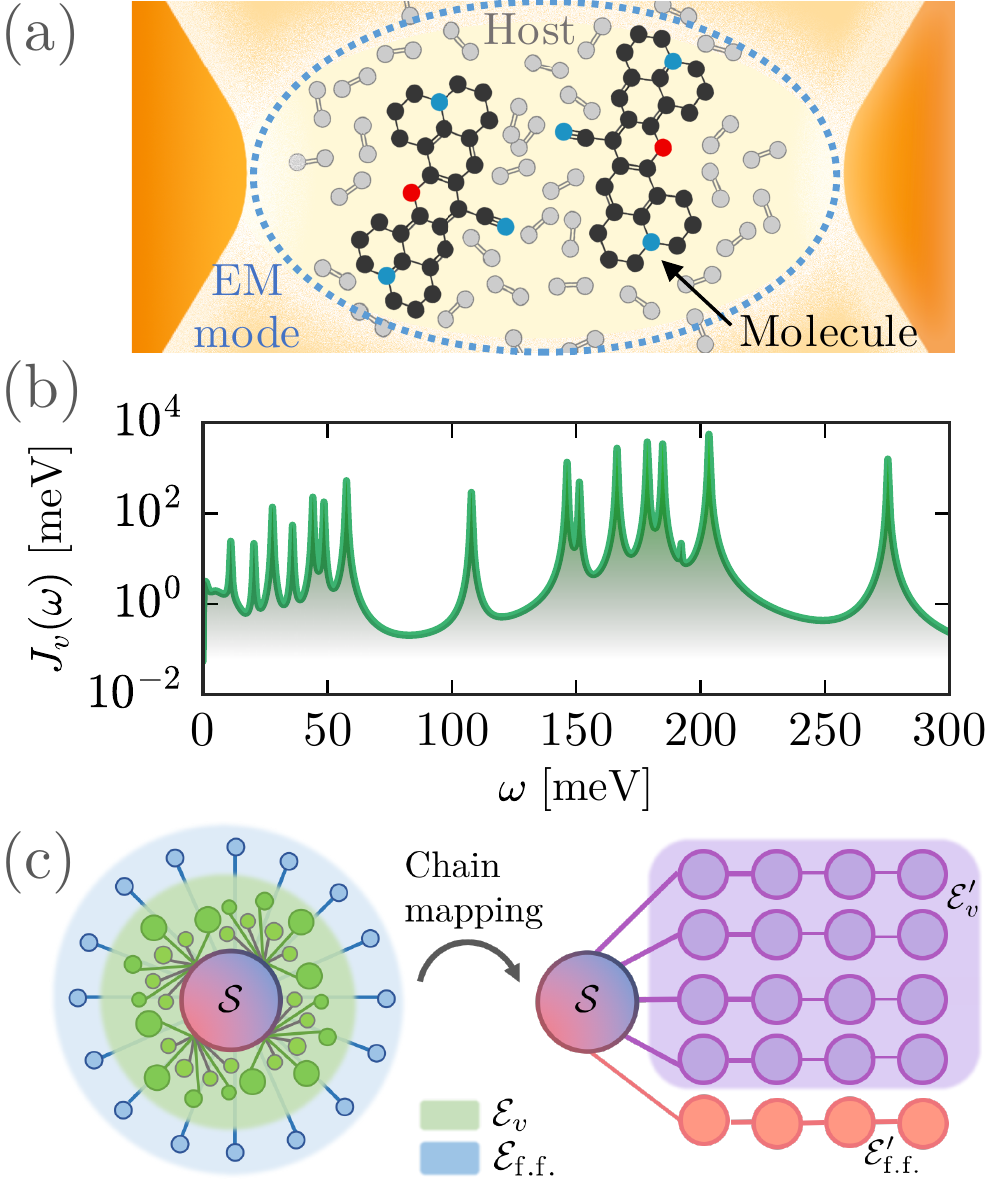}
		\caption{(a) Sketch of a molecular ensemble interacting with a confined EM
			resonance (dashed region) and with the host environment (grey circles). (b)
			Vibrational spectral density for Rhodamine 800 molecule.  (c) Scheme
			illustrating the mapping of the vibrational modes yielding the multi-chain
			Hamiltonian used in the simulations.}
		\label{fig:setup}
	\end{figure}
	
	In this Letter we simulate the temporal dynamics of the polaritons supported by
	an optical microcavity filled with an ensemble of prototypical organic molecules
	(Rhodamine 800), as sketched in \autoref{fig:setup}a. In our approach, nuclear,
	electronic and photonic processes are treated fully and on equal footing by
	employing the Time-Dependent Variational Matrix Product States algorithm
	(TDVMPS), which can provide quasi-exact solutions of open quantum system
	dynamics by truncating the maximum entanglement between different system
	components~\cite{Prior2010, Haegeman2011, Chin2013, Wall2016Simulating,
		Schroder2016Simulating}. Our fundamental study reveals a variety of regimes for
	the time evolution of the polariton populations, which experience
	vibration-assisted dynamics that translate into distinct far-field emission
	fingerprints. We observe clear signatures of non-Markovian behavior in the whole
	range of light-matter coupling, both for single and multiple molecules coupled
	to an EM cavity mode. We find that the full vibrational structure of the
	molecules has to be treated for a complete understanding of the interplay
	between excitons, phonons, and photons.
	
	Our model includes a collection of $N$ identical molecules with two
	electronic states (frequency $\omega_e$ and ladder operators
	$\hat{\sigma}^{(i)}_\pm$, $i\in [1,\ldots,N]$), placed within the volume of a
	nano- or microcavity supporting a single dispersionless EM
	mode (frequency $\omega_O=\omega_e=2.5$~eV), with annihilation operator
	$\hat{a}$. The total Hamiltonian contains three different parts, as
	schematically depicted in \autoref{fig:setup}c. First, the ``bare'' system
	$\mathcal{S}$ accounts for the excitons within the molecules, the cavity EM
	mode, and their coupling, measured by the collective Rabi frequency
	$\Omega_R$ and treated within the rotating wave approximation (setting
	$\hbar=1$),
	\begin{equation}
		\hat{H}_{\mathcal{S}}= \omega_O\hat{a}^{\dagger}\hat{a}+\sum_{i=1}^{N}
		\omega_e \hat{\sigma}^{(i)}_{+}\hat{\sigma}^{(i)}_{-} + \frac{\Omega_R}{2\sqrt{N}}
		\sum_{i=1}^{N}(\hat{a}^{\dagger}\hat{\sigma}^{(i)}_-+\hat{\sigma}^{(i)}_+\hat{a}) \label{eq:Our_H}.
	\end{equation}
	Direct dipole-dipole interactions at typical intermolecular
	distances in strong-coupling experiments only provide small
	corrections~\cite{Cwik2014, Gonzalez-Ballestero2016Uncoupled,
		Saez-Blazquez2017} and are thus neglected here for simplicity and
	generality. In the single-excitation subspace, $\hat{H}_{\mathcal{S}}$ is
	exactly solvable, with two types of eigenstates: (i) upper (UP) and lower
	(LP) polaritons $\ket{\pm}= (\hat{a}^{\dagger}\ket{G} \pm \ket{B}) /
	\sqrt{2}$ with frequencies $\omega_{\pm}=\omega_O\pm \Omega_R/2$, which
	result from the hybridization of the collective excitonic \emph{bright}
	state $\ket{B}=(\sum_{i=1}^N \hat{\sigma}_+^{(i)}\ket{G})/\sqrt{N}$ with the
	cavity EM mode (with $\ket{G}$ the global vacuum state). (ii) $N\!-\!1$
	purely excitonic \emph{dark} states (DS), $\ket{d}\in \mathcal{D}$,
	orthogonal to $\ket{B}$, with frequency $\omega_e$.
	
	The second part of the Hamiltonian describes the vibrational subspace
	$\mathcal{E}^{(i)}_v$ containing $M_v$ vibrational modes for each molecule, and
	their elastic coupling to the excitons. The $k$th vibrational mode is
	approximated by a harmonic oscillator of frequency $\omega_k$ (valid close to
	the equilibrium position) with annihilation operator $\hat{b}^{(i)}_{k}$ and
	$k$-dependent exciton-phonon coupling strength $\lambda^{(i)}_{k}$,
	\begin{equation}
		\hat{H}_{v} = \sum_{i=1}^{N} \sum_{k=1}^{M_v} \omega_{k} \hat{b}_{k}^{\dagger (i)}\hat{b}^{(i)}_{k} \, +
		\sum_{i=1}^{N} \sum_{k=1}^{M_v} \lambda^{(i)}_{k} (\hat{b}^{(i)}_{k}+\hat{b}_{k}^{\dagger (i)})
		\hat{\sigma}^{(i)}_{+}\hat{\sigma}^{(i)}_{-} .\label{eq:mol_H}
	\end{equation}
	The properties of these modes, $\{\omega_k,\lambda_k^{(i)}\}$, are encoded
	in the spectral density, assumed to be identical for all molecules,
	$J_v^{(i)}(\omega) = J_v(\omega) = \pi\sum_{k=1}^{M_v}
	\lambda_k^2\delta(\omega-\omega_k)$. For Rhodamine 800 this density is
	extracted from the spectroscopic measurements in
	Ref.~\cite{Christensson2010}, as displayed in \autoref{fig:setup}b, with
	vibrational frequencies located in the range $[0,0.3]$~eV, and
	reorganization energy $\Delta = \int_0^{\infty}
	\frac{J_v(\omega)}{\pi\omega} \mathrm{d}\omega \approx 35.6$~meV. While no
	interactions between vibrational modes are included in the model, internal
	vibrational decay due to interactions with the host medium is partially
	represented through the non-zero width of the peaks in $J_v(\omega)$.
	Furthermore, we have checked that the results presented below do not depend
	sensitively on the properties of $J_v(\omega)$.
	
	The third part of the Hamiltonian describes radiative far-field photon modes
	$\hat{f}_l$ and their coupling to the cavity EM mode, $\hat{H}_{r} = \sum_{l}
	\omega_{l} \hat{f}_l^\dagger \hat{f}_l + \eta_l
	(\hat{a}^{\dagger}\hat{f}_l+\hat{f}_l^{\dagger}\hat{a})$~\footnote{Note that the
		coherent coupling to free-space radiative modes induces a small Lamb shift in
		the effective cavity frequency that depends on the numerical cutoff
		$\omega_{\mathrm{r}}^{\mathrm{cut}}$ ($=3.5$~eV in this work).}. In a way
	similar to vibrational modes, we introduce here the spectral density for the
	photonic subspace, $\mathcal{E}_\mathrm{r}$, as $J_{\mathrm{r}}(\omega) =
	\pi\sum_{l}\eta_l^2 \delta(\omega-\omega_l) = \kappa\omega^3/(2\omega_O^3)$. We
	set the bare-cavity decay rate $\kappa=2J_{\mathrm{r}}(\omega_O)$ to $50$~meV,
	typical for plasmonic/dielectric cavities.
	
	For typical molecules, the large number of vibrational modes, $M_v\sim 10^2$,
	makes direct diagonalization of the total Hamiltonian infeasible. We resolve
	this by applying the TDVMPS approach. We first perform an orthogonal chain
	mapping of the modes in the $N$ vibrational (green) and the free-space photon
	(blue) environments ($\mathcal{E}^{(i)}_\mathrm{v}$ and
	$\mathcal{E}_{\mathrm{r}}$), sketched in \autoref{fig:setup}c and detailed
	in~\cite{supplemental}, regrouping them in chains with length $L=M_v$ with
	nearest-neighbor hopping, with only the first chain mode coupled to the
	exciton-photon subspace $\mathcal{S}$ (red-blue)~\cite{Chin2011}. The wave
	function $\ket{\psi(t)}$ is represented by a tensor
	network~\cite{Schollwock2011} with maximum bond dimensions $D$ with a structure
	mimicking the transformed Hamiltonian. If a single root tensor stores the system
	$\mathcal{S}$, its size scales exponentially with $N$, leading to a severe
	memory bottleneck. This scaling can be efficiently reduced while maintaining
	precision by decomposing $\ket{\psi(t)}$ into a \emph{tree} tensor network, with
	a structure determined to minimize the entanglement between
	nodes~\cite{Szalay2015, Shi2006}, and each final branch coupled to a single
	chain (see~\cite{supplemental} for details). This allows for the treatment of
	$N=16$ molecules coupled to $N+1$ environments with $L=350$ modes each, i.e., a
	system for which the full Hilbert space has a dimension of $\approx 50^{(N+1)
		L}$ (allowing $50$ basis states per phonon mode), with $(N+1) L = 5950$, through
	a wavefunction described by $\approx 10^8$ parameters. To further ameliorate
	memory issues for large chain mode occupations, we employ an optimal boson basis
	for the chain tensors, determined on the fly~\cite{Guo2012}. A more detailed
	description of the theoretical approach can be found
	in~\cite{Schroder2017Multi,DelPino2018Ground}. We here focus on the time
	evolution after excitation, but note that the same approach also allows
	efficient calculation of the ``lower
	polaron-polariton''~\cite{Spano2015,Herrera2016,Wu2016,Zeb2018}, as shown
	in~\cite{DelPino2018Ground}.
	
	\begin{figure}
		\includegraphics[width=\linewidth]{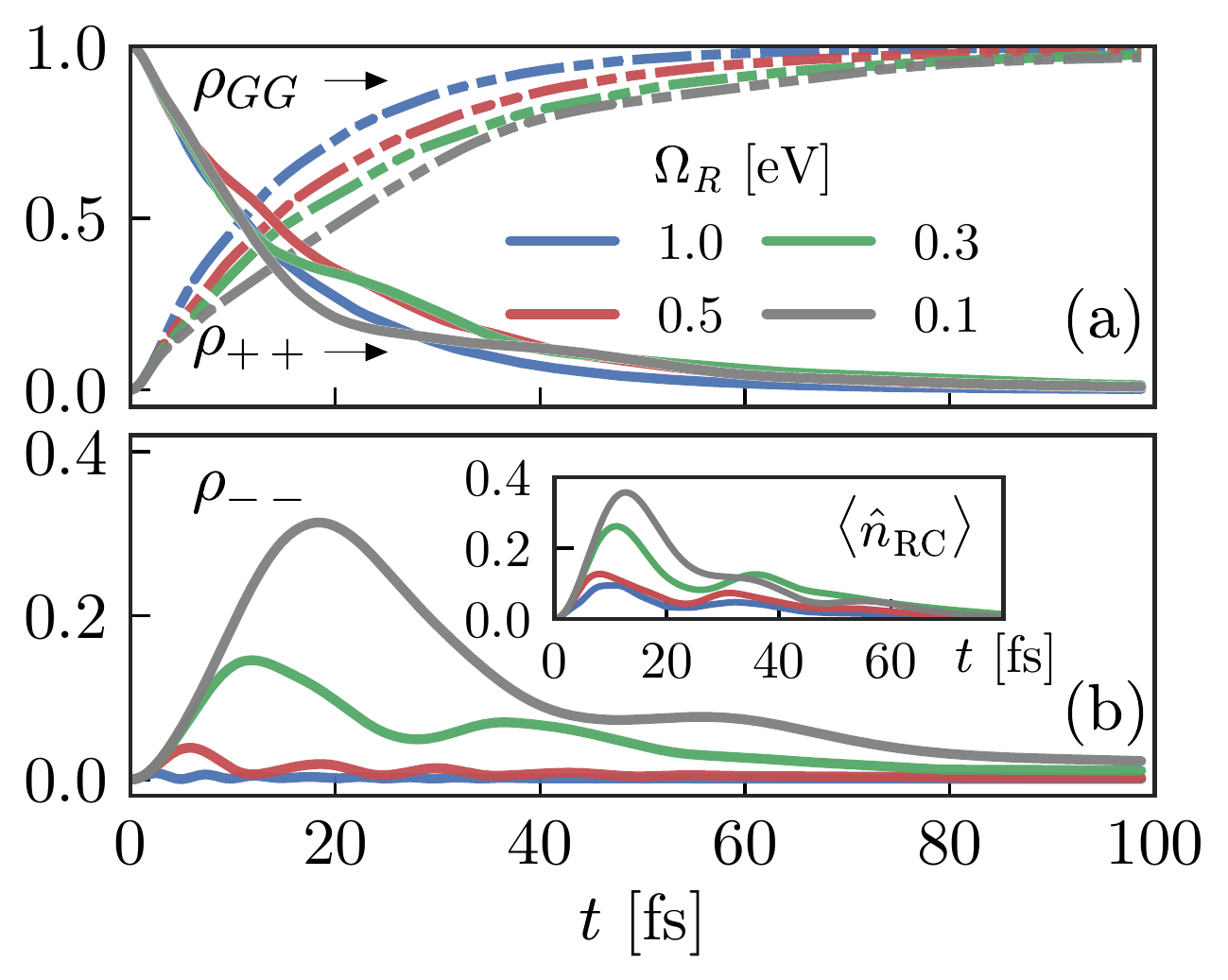}
		\caption{Dynamics of reduced density matrix populations for a single
			molecule as a function of the Rabi frequency $\Omega_R$, including (a)
			vacuum and upper polariton states and (b) the lower polariton. The inset in
			(b) shows the occupation of the collective reaction coordinate in the
			vibrational environment.}
		\label{fig:fig2}
	\end{figure}
	
	\begin{figure*}
		\includegraphics[width=\linewidth]{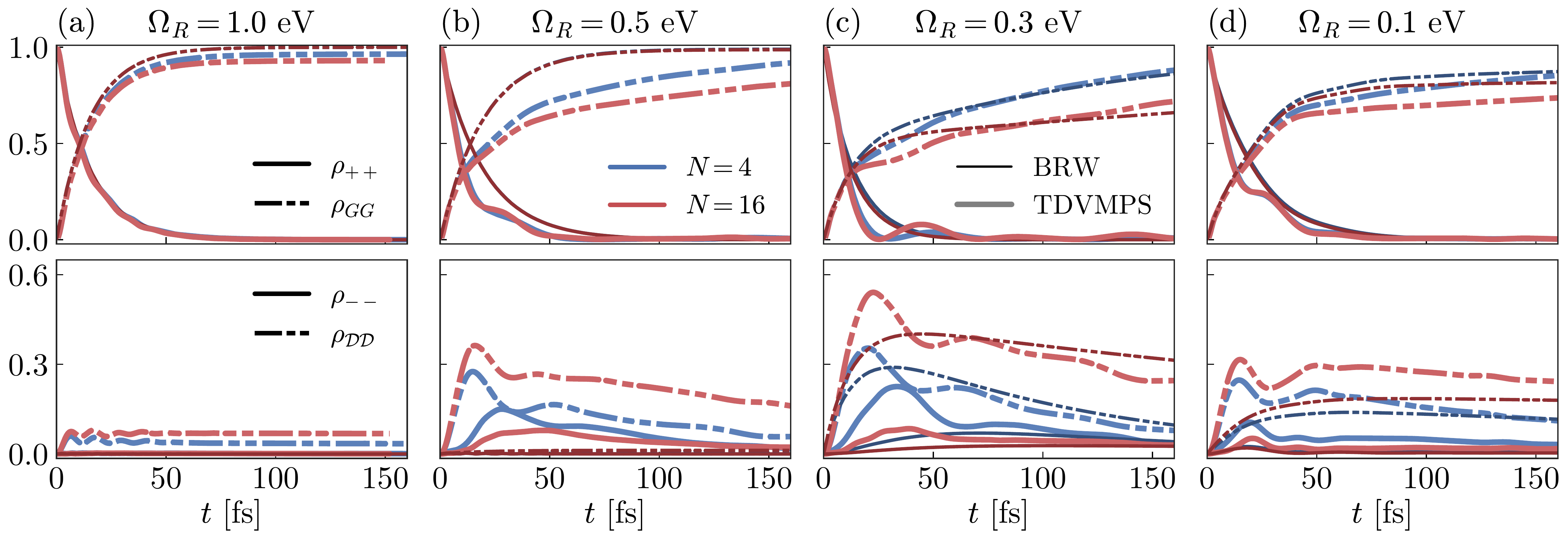}
		\caption{Population dynamics for $N=16$ at different Rabi frequencies (shown
			in titles). The occupations $\rho_{GG}$, $\rho_{++}$ are displayed in the
			upper panels, while $\rho_{--}$, $\rho_{\mathcal{D}\mathcal{D}}$ is shown in
			the lower ones, with distinctive line styles. Different colours depict the full numerical TDVMPS result and the Markovian limit calculated as described in the text.}
		\label{fig:fig3}
	\end{figure*}
	
	We first explore single-molecule strong coupling ($N=1$, for which there are no
	dark states), recently realized in plasmonic nanocavities~\cite{Zengin2013,
		Chikkaraddy2016}. We set the initial state ($t=0$) to the bare (vibrationally
	undressed) UP, $\ket{+}$, as would be produced by an on-resonance laser pulse
	short enough to ensure that nuclear motion can be neglected during its action.
	This allows to restrict the simulation to only the zero- and single-excitation
	subspaces of $\hat H_{\mathcal{S}}$. \autoref{fig:fig2} shows the time evolution
	of the populations $\{\rho_{GG},\rho_{\pm\pm}\}$, extracted from the system
	density matrix $\hat{\rho}_{\mathcal{S}}(t) =
	\mathrm{Tr}_{\mathcal{E}_\mathrm{v},\mathcal{E}_{\mathrm{r}}}
	\{\ket{\psi(t)}\bra{\psi(t)}\}$. Details of the numerical parameters and
	convergence behavior can be found in the supplemental
	material~\cite{supplemental}. When $\Omega_R$ ($=1$~eV in \autoref{fig:fig2}a)
	is much larger than the vibrational frequencies
	($\omega_\mathrm{v}^\mathrm{cut}=0.3$~eV), the bare UP ($\rho_{++}$) shows rapid
	exponential decay to the ground state ($\rho_{\mathrm{GG}}$) through photon
	emission, with a negligible population increase of the LP. This is exactly what
	an approach treating $\mathcal{E}_\mathrm{v}$ and $\mathcal{E}_{\mathrm{r}}$ as
	Markovian (memory-less) baths would predict, as there are no available phonon
	modes that could induce transitions between UP and LP\@. However, already for
	the case $\Omega_R=0.5$~eV, our results show a new decay pathway for the UP,
	which can relax through phonon emission to the LP\@. A closer look reveals that
	the emission of photons, $\rho_{++}\rightarrow\rho_{GG}$, is
	superimposed to a coherent exchange of population between UP and LP (see
	\autoref{fig:fig2}b). Moreover, this dynamics is accompanied by a collective
	excitation of the vibrational modes, as displayed by the time evolution of
	the reaction coordinate population, $\hat{b}_{\mathrm{RC}} =
	\sum_k\lambda_k\hat{b}_k/\sqrt{\sum_k\lambda_k^2}$ (inset of
	\autoref{fig:fig2}b). Neither of these effects could be reproduced by a
	Markovian approximation, which would lead to single-phonon transitions between
	polaritons at rates $\Gamma^v_{if}=2J_v(\omega_{if})$, where
	$\omega_{if}=\omega_i-\omega_f$ is the transition frequency. As the Rabi
	splitting is larger than the largest vibrational frequency available, decay into
	the LP is forbidden within the Markovian limit~\cite{Breuer2007}. The amplitude
	of these oscillations is enhanced for lower Rabi frequencies, where vibronic and
	photonic couplings become comparable (see cases $\Omega_R=0.3$~eV and
	$\Omega_R=0.1$~eV in \autoref{fig:fig2}b). Here the onset of non-exponential
	behavior in both $\rho_{++}$ and $\rho_{GG}$ is noticed, resulting in delayed
	photon emission. In particular, while the UP starts emitting photons
	immediately, the LP population is maximized after approximately one cycle of
	coherent oscillation of the reaction coordinate before radiative decay starts to
	dominate.
	
	We next proceed to discuss the many-molecule case, in which dark states have a
	severe impact on the dynamics~\cite{Litinskaya2006}. In this section we compare
	our results with those emerging from a standard master equation derived using
	the Markovian Bloch-Redfield-Wangsness (BRW)
	approach~\cite{Wangsness1953,Redfield1955}, which considers solely the value of
	$J_v(\omega)$ at the transition frequency within the so-called secular
	approximation and is restricted to single-phonon transitions. We note that while
	it is possible to derive more advanced Markovian and non-Markovian master
	equations~\cite{Wilson-Rae2002,McCutcheon2010,Roy2011}, BRW theory already goes
	significantly beyond the widely employed Lindblad master equation approach, and
	allows clear identification of non-Markovian and multi-phonon processes in the
	TDVMPS simulations.
	
	For very large $\Omega_R$ (see \autoref{fig:fig3}a), the UP decays mostly by
	photon emission. While vibrational decay to the LP is negligible, some
	population does reach the DS (with $\rho_{\mathcal{D}\mathcal{D}}=\sum_{d}
	\rho_{dd}$). Remarkably, while the photonic decay can be accurately determined
	by BRW theory, the prediction for the DS pathway disagrees with the TDVMPS
	calculation. BRW theory predicts $\rho_{\mathcal{D}\mathcal{D}}=\rho_{--}=0$ as
	there are no phonon modes at the required transition frequencies, while TDVMPS
	shows that the DS reservoir is indeed populated. This demonstrates that
	vibration-driven decay from polaritonic
	states~\cite{Litinskaya2004,Coles2014,DelPino2015Quantum,Neuman2018} can occur
	efficiently even when no vibrational modes are resonant with the transition
	frequency, and that the decoupling from vibrational modes that is found for the
	LP under collective strong coupling~\cite{Galego2015, Herrera2016, Galego2016,
		Zeb2018, DelPino2018Ground} does not prevent decay of the UP\@.
	
	For smaller values of the Rabi splitting, the UP-DS and DS-LP transition
	frequencies lie within the range of $J_v(\omega)$. \autoref{fig:fig3}b shows
	that for $\Omega_R=0.5$~eV, coherent population transfer to the DS competes with
	the fast photonic decay of the UP, inducing threefold oscillations $\rho_{++}
	\leftrightarrow \rho_{\mathcal{D}\mathcal{D}} \leftrightarrow \rho_{--}$ that
	persist over more than $50$~fs. For a larger number of molecules, population is
	``trapped'' more efficiently in $\mathcal{D}$ and subsequently decays to the LP
	over several hundred fs before being emitted (see~\cite{supplemental} for a
	comparison with the case $N=4$). While the intrinsic lifetimes of the LP and UP
	are similar due to efficient photon leakage out of the cavity, as seen in
	simulations initialized in the LP (not shown), the refilling from the DS leads
	to its population persisting over much longer timescales. This observation
	agrees with the long-time tails observed in strongly coupled
	J-aggregates~\cite{Schwartz2013, George2015Ultra-strong}. As the UP-DS-LP
	transition frequencies are not resonant with any vibration in the system
	(cf.~\autoref{fig:setup}b) the DS and LP remain unpopulated within BRW theory.
	In contrast, TDVMPS can represent both multi-phonon relaxation and the
	broadening of the polaritons due to decay, reducing the stringency of
	vibrational resonance conditions.
	
	Similarly to the single-molecule case, vibrationally-driven oscillations take
	longer to relax for smaller Rabi frequencies. Interestingly, the numerical
	agreement between Markovian and non-Markovian approaches is improved in the
	particular case of $\Omega_R=0.3$~eV (\autoref{fig:fig3}c), where the UP-DS
	transition is quasi-resonant with a vibrational resonance (cf.
	\autoref{fig:setup}c). In this case, while the long-time behavior is reasonably
	well approximated by BRW theory, the rapid short-time oscillatory dynamics is
	averaged out. For even smaller Rabi frequencies, we do not observe a monotonic
	increase of the LP population, as opposed to the case $N=1$. In particular,
	\autoref{fig:fig3}d displays the case $\Omega_R=0.1$~eV, where the UP population
	is only slightly more efficiently transferred to DS and LP than for
	$\Omega_R=0.5$~eV (\autoref{fig:fig3}b). Here the large phonon coupling leads to
	a rapid destruction of polariton coherence, i.e., loss of strong
	coupling~\cite{Torma2015}.
	
	In the supplemental material~\cite{supplemental}, we additionally compare TDVMPS
	with the single-mode HTC model, which has been successfully used to predict
	non-Markovian dynamics and energy transfer between exciton-polaritons and
	DS~\cite{Spano2015, Herrera2016, Herrera2017Dark, Herrera2017Absorption,
		Herrera2017Theory}. While it reproduces the dynamics in the first few fs
	(dominated by the reaction coordinate response) reasonably well, it consistently
	overestimates the coherent oscillations observed for $t\gtrsim 10$~fs. In
	particular, it fails to correctly predict the excitation trapping in
	$\mathcal{D}$ and instead leads to enduring oscillations.
	
	\begin{figure}
		\includegraphics[width=\linewidth]{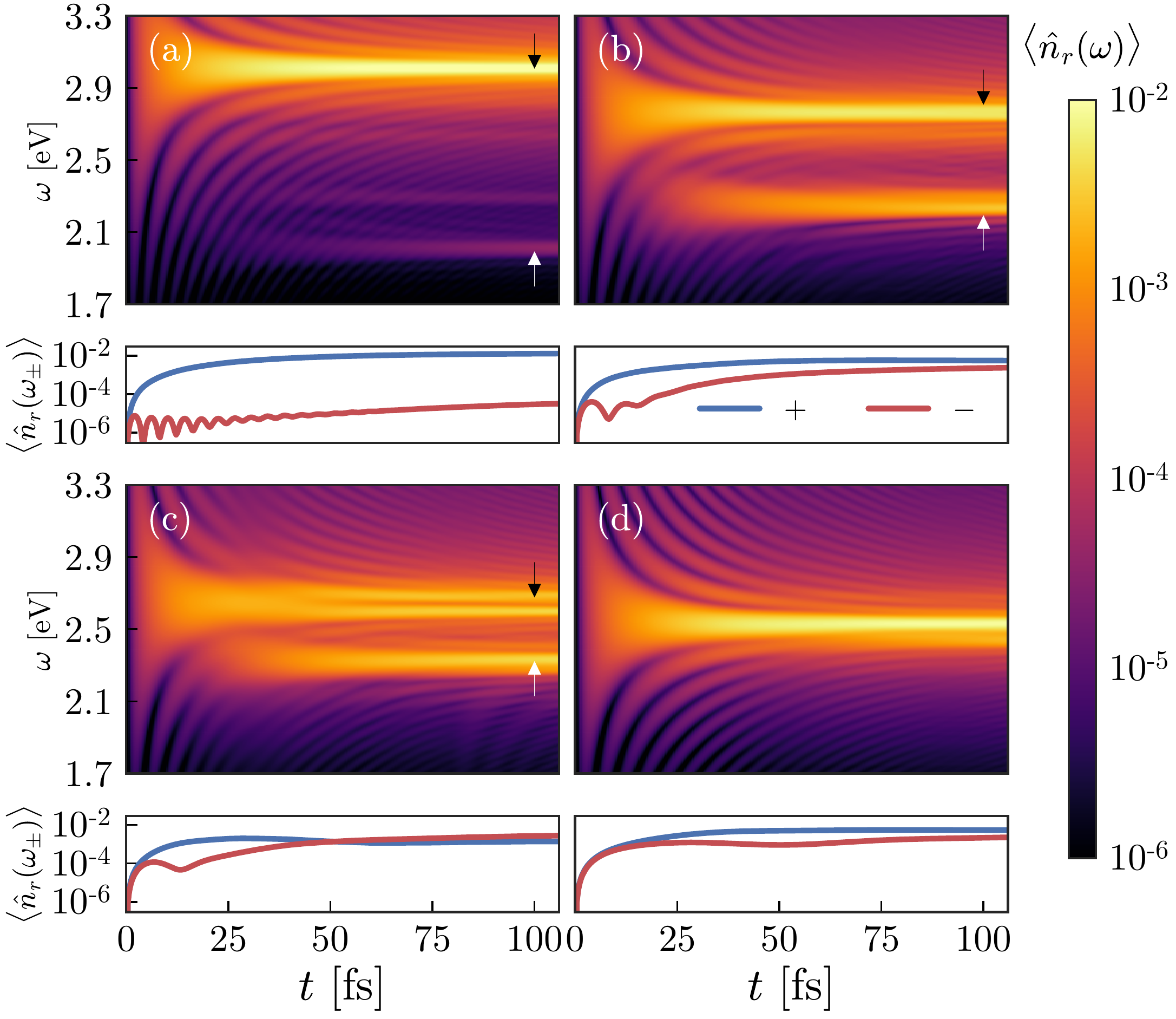}
		\caption{Far-field photonic population as a function of frequency and time,
			for the same Rabi splittings as in \autoref{fig:fig3} and $N=4$. The black and white
			arrows in panel (a-c) point to the approximate location of the bare UP and
			LP. Each panel includes in the lower parts cuts at $\omega_\pm$.}
		\label{fig:fig4}
	\end{figure}
	
	In typical organic polariton experiments, the collected far-field photons are
	the main source of available information. Specifically, short-time energy
	redistribution between polaritons could be traced by ultrafast pump-probe
	experiments \cite{Savvidis2006,Song2004,Berera2009,Schwartz2013}. The final part
	of this work is thus devoted to the fingerprints displayed in the time-resolved
	emission spectrum arising from the different dynamics regimes that have been
	analyzed above. Thanks to the full access to the reservoir degrees of freedom
	provided by TDVMPS~\cite{Schroder2016Simulating}, emission can be retrieved from
	the occupation of the far-field modes $\braket{\hat{n}_{\mathrm{r}}(\omega_l,t)}
	= \braket{\psi(t)|\hat{f}_l^{\dagger}\hat{f}_l|\psi(t)}$.
	
	For very large Rabi splittings ($\Omega_R=1$~eV), \autoref{fig:fig4}a shows
	dominant photonic emission from the UP, with a series of side lobes converging
	to the main emission line due to the coherent buildup of population in the
	free-space mode~\cite{Argenti2013}. At timescales comparable to the reaction
	coordinate dynamics ($\tau_{\mathrm{RC}}=2\pi/\omega_{\mathrm{RC}}\simeq23$ fs),
	population transfer through the DS reservoir reaches the LP and its emission is
	observed as well, as clearly seen in cuts at the bare polariton frequencies
	$\omega_{\pm}$ (lower panel of \autoref{fig:fig4}a) that display the buildup of
	far-field occupation. The asymptotic growth of the photons emitted by the LP
	continues after the UP is practically depopulated, due to the continued
	refilling from the dark states. In addition to the vibration-free polaritons
	$\ket{\pm}$, emission bands at intermediate energies are visible in the
	spectrum. This is interpreted as due to small
	cavity admixtures to (vibrationally relaxed) dark states, in line with
	experimental observations~\cite{George2015Ultra-strong,Baieva2017}.
	
	As the Rabi frequency is diminished, more efficient coupling from UP to DS
	increases population transfer and induces brighter emission from the LP and its
	vibronic sidebands. For $\Omega_R=0.3$~eV, depicted in \autoref{fig:fig4}c, the
	UP-DS and DS-LP transitions are close to resonant with a vibrational mode, and
	splitting of the emission from the UP is observed, consistent with the coherent
	oscillations in \autoref{fig:fig3}c. This is reminiscent of the splitting
	between vibrationally dressed and undressed polaritonic states found in the HTC
	model~\cite{Herrera2017Theory,Herrera2017Dark,Herrera2017Absorption}. Finally,
	for $\Omega_R=0.1$~eV (\autoref{fig:fig4}d), no Rabi splitting is observed,
	suggesting that the system is in the weak-coupling regime where no polaritons
	are formed.
	
	To conclude, we have unveiled the temporal dynamics of organic polaritons. We
	have employed a powerful quasi-exact tree tensor network algorithm, which has
	enabled us to treat a highly structured reservoir of molecular vibrations and
	free-space emission of the cavity without additional approximations. The
	simulations reveal coherent vibration-driven oscillations between polaritons and
	dark states that are only weakly dependent on resonance conditions due to the
	strong exciton-phonon coupling. This demonstrates the importance of multi-phonon
	processes and non-Markovian dynamics in the system, which are easily
	underestimated or overestimated in simplified frameworks such as the
	Bloch-Redfield-Wangsness approximation or the Holstein-Tavis-Cummings model.  In
	addition, the time-resolved emission spectra show fast energy relaxation to the
	lower polariton, on the scale of tens of femtoseconds. In contrast to Kasha's
	rule for bare molecules, the radiative and vibrational decays of the upper
	polariton are similarly fast, and its emission is clearly observed.
	
	
	\begin{acknowledgments}
		This work has been funded by the European Research Council (ERC-2011-AdG-290981
		and ERC-2016-STG-714870), by the European Union Seventh Framework Programme
		under grant agreement FP7-PEOPLE-2013-CIG-618229, and the Spanish MINECO under
		contract MAT2014-53432-C5-5-R and the ``María de Maeztu'' programme for Units of
		Excellence in R\&D (MDM-2014-0377). F.A.Y.N.S. and A.W.C. gratefully acknowledge
		the support of the Winton Programme for the Physics of Sustainability and
		EPSRC\@.
	\end{acknowledgments}
	
	\bibliography{references,extranotes}

\clearpage	
\pagebreak
\renewcommand{\theequation}{A.\arabic{equation}}
\onecolumngrid
\begin{center}
	\textbf{\large Supplemental Material}
\end{center}
\setcounter{equation}{0}
\setcounter{figure}{0}
\setcounter{table}{0}
\setcounter{page}{1}
\makeatletter
\vspace{0.9cm}
\twocolumngrid

	\section{Chain mapping}

\begin{figure}[tb]
	\centering
	\includegraphics[width=1\linewidth]{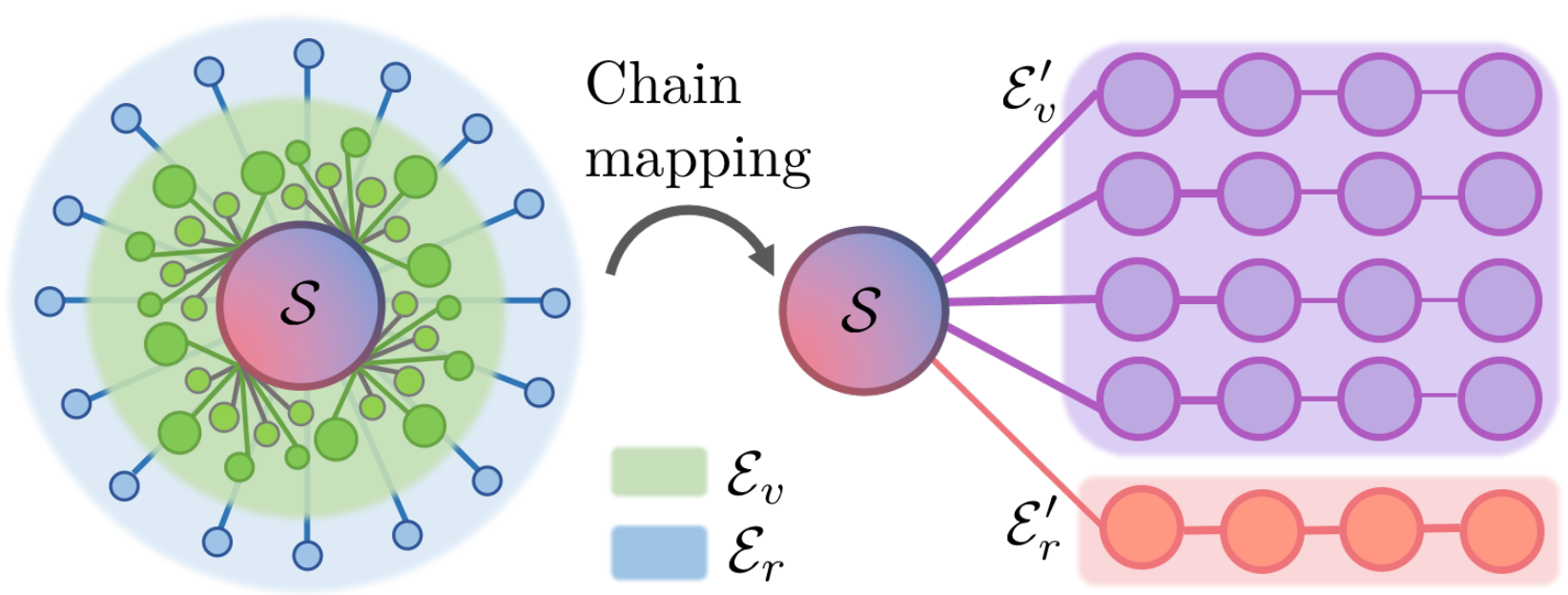}
	\caption{Chain mapping of the multi-environment Hamiltonian in the main text to a star-structure that includes excitonic and photonic degrees of freedom in the root operator.}\label{fig:chain_mapping_2}
\end{figure}

We here give a short overview of the chain mapping that converts an environment
of multiple bosonic modes with linear coupling to a system into a chain of
bosonic modes with nearest-neighbor hopping, with only the first chain site
coupled to the system~\cite{Chin2011s}. Such an environment is described by
the general Hamiltonian
\begin{align}
	\hat{H}_{\mathcal{E}} &= \sum_{k=1}^{M} \left[ \omega_{k} \hat{b}_{k}^\dagger \hat{b}_{k}^{\phantom{\dagger}} +
	(\lambda_{k} \hat{O}_\mathcal{S}^\dagger \hat{b}_{k} + \mathrm{H.c.}) \right] ,
\end{align}
where $\hat{O}_\mathcal{S}$ is an arbitrary system operator, and $\hat{b}_k$ the
bath oscillator annihilation operators. The environment is fully characterized
through the spectral density $J(\omega) = \pi \sum_k \lambda_k^2
\delta(\omega-\omega_k)$. Rewriting $\hat{H}_{\mathcal{E}} =
\hat{\boldsymbol{\beta}}^{\dagger} \mathcal{C} \hat{\boldsymbol{\beta}}$, where
$\hat{\boldsymbol{\beta}} = (\hat{O}_\mathcal{S}, \hat{b}_{1}, \cdots
\hat{b}_{M})^{T}$, the chain mapping is obtained by tridiagonalization (e.g.,
with the Lanczos algorithm~\cite{Lanczos1950s}) of the coefficient matrix
$\mathcal{C} = \left(\begin{smallmatrix} 0 & \boldsymbol{\lambda} \\
\boldsymbol{\lambda}^\dagger
&\bar{\boldsymbol{\omega}}\end{smallmatrix}\right)$, where
$\boldsymbol{\lambda}=(\lambda_1,\cdots,\lambda_{M})$ and
$\bar{\boldsymbol{\omega}}=\mathrm{diag}(\omega_1,\cdots,\omega_{M})$. This
gives $\hat{H}_{\mathcal{E}} = \hat{\boldsymbol{\beta}}'^{\dagger}
\mathcal{C} \hat{\boldsymbol{\beta}}'$, with $\hat{\boldsymbol{\beta}}' =
(\hat{O}_\mathcal{S}, \hat{c}_{1}, \cdots \hat{c}_{M})^{T}$, and
\begin{align}
	C' = U^\dagger C U =
	\begin{pmatrix}
		0 & \eta & 0\\
		\eta & \tilde{\omega}_1 & t_{1}\\
		0 & t_{1} & \tilde{\omega}_{2} & \ddots\\
		&  & \ddots & \ddots & t_{M-1}\\
		&  &  & t_{M-1} & \tilde{\omega}_{M}
	\end{pmatrix},
\end{align}
i.e., the desired chain Hamiltonian,
\begin{multline}
	\hat{H}_{\mathcal{E}} = \sum_{k=1}^M \tilde{\omega}_k\hat{c}_k^{\dagger} \hat{c}_k
	+ \eta \left(\hat{O}_\mathcal{S} \hat{c}_{1}^\dagger + \hat{O}_\mathcal{S}^\dagger \hat{c}_{1}\right)\\
	+ \sum_{k=1}^{M-1} t_k \left(\hat{c}^{\dagger}_{k}\hat{c}_{k+1}+\hat{c}^{\dagger}_{k+1}\hat{c}_{k}\right),
	\label{eq:H_chain_general}
\end{multline}
where the reaction coordinate that interacts with the system is given by
$\hat{c}_1 = \sum_k\lambda_k\hat{b}_k/\eta$, with coupling $\eta = \sqrt{\sum_k
	|\lambda_k|^2}$ and frequency $\tilde{\omega}_1 = \sum_k \omega_k |\lambda_k|^2
/ \eta^2$.

\section{Tree tensor network}

For the system treated in the main text (a collection of Rh800 molecules
coupled to a cavity mode), the phononic bath $\mathcal{E}_\mathrm{v}^{(i)}$ on
molecule $i$ interacts with the molecular exciton through the operator
$\hat{O}_\mathcal{S} = \sigma^{(i)}_+\sigma^{(i)}_-$, while the free-space
photon modes $\mathcal{E}_\mathrm{r}$ interact with the cavity photon via
$\hat{O}_\mathcal{S} = a$ (in the rotating-wave approximation). For the
photons, the spectral density $J_r(\omega)\propto\omega^3$ is of the Leggett
form ($\propto \omega^s$, $s>0$)~\cite{Leggett1987s}, which enables closed
expressions for all $\{\tilde\omega_k,t_k\}$~\cite{Prior2010s,Chin2011s}.

\begin{figure}[tb]
	\centering
	\includegraphics[width=1\linewidth]{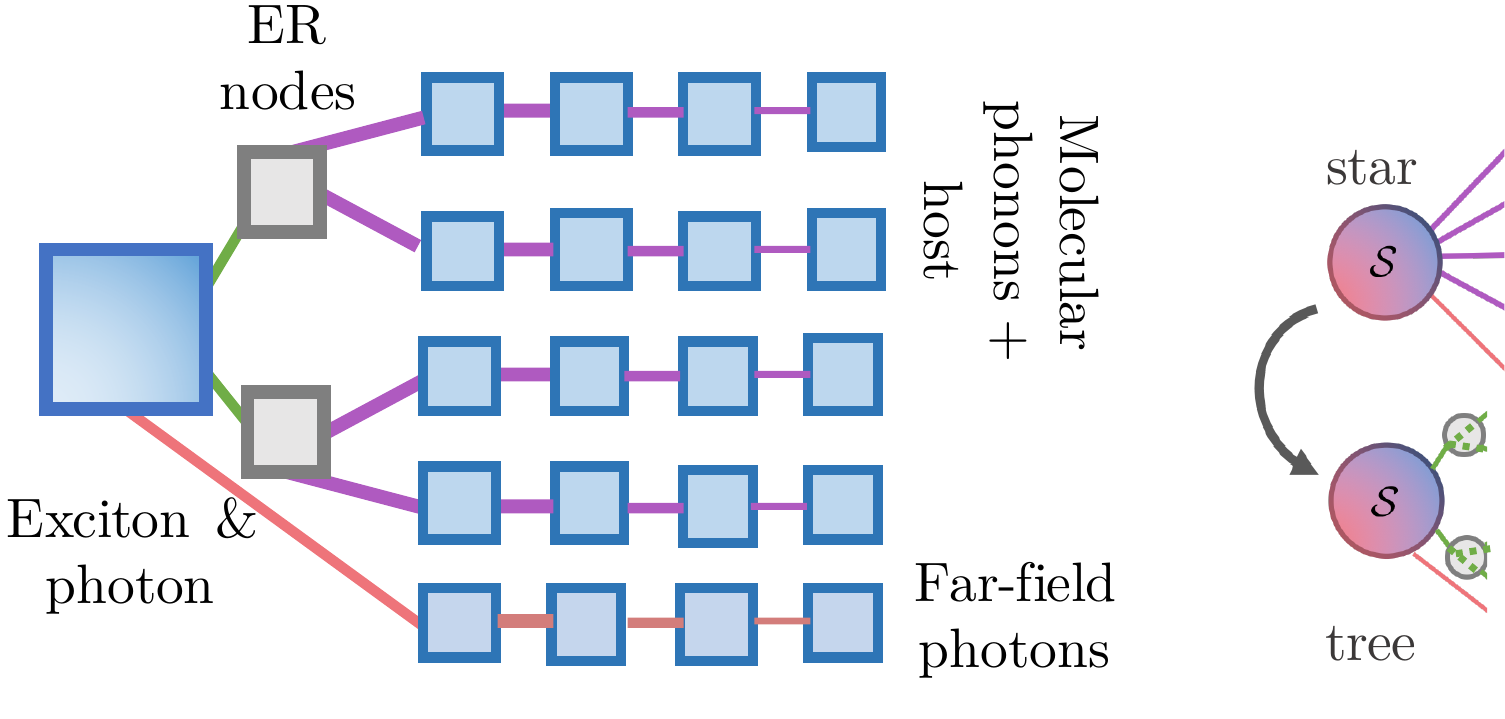}
	\caption{Representation as a tree-like tensor network by further singular-value decompositions of the root node in the star-tensor-network that mimicks \autoref{fig:chain_mapping_2} (left). This requires to introduce auxiliary non-physical sites, where the acting Hamiltonian (right) presents empty operators (grey circles).}\label{fig:chain_mapping_tree}
\end{figure}

\begin{figure*}[tb]
	\centering
	\includegraphics[width=1\linewidth]{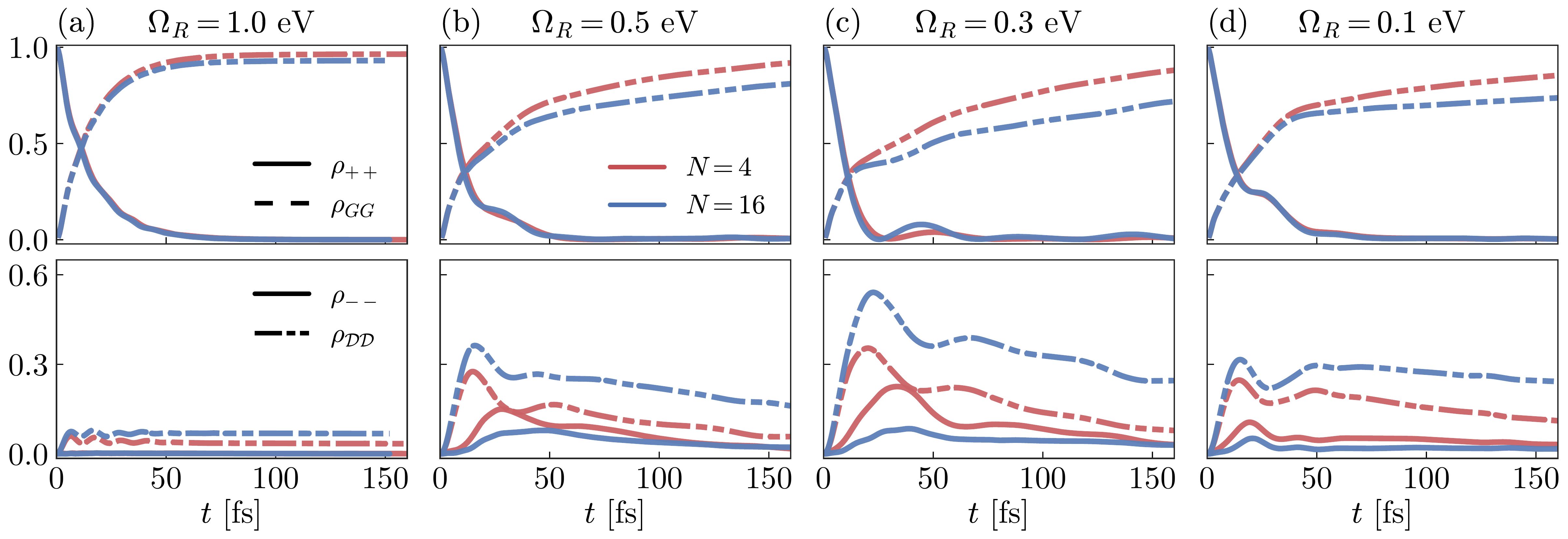}
	\caption{Population dynamics for $N=4$ (red) and $N=16$ (blue) molecules for
		different Rabi frequencies (shown in titles). The occupations
		$\rho_{GG}$, $\rho_{++}$ are displayed in the upper panels, while
		$\rho_{--}$, $\rho_{\mathcal{D}\mathcal{D}}$ is shown in the lower ones,
		with distinctive line styles.}\label{fig:N4_vs_N16}
\end{figure*}

The application of the chain mapping to all environments yields the ``star''
Hamiltonian sketched in \autoref{fig:chain_mapping_2}. This Hamiltonian only
contains nearest-neighbor coupling terms and thus can be efficiently
implemented in tensor-network descriptions that share the same network
topology~\cite{Schollwock2011s,Schroder2017Multis}. Since the chain mapping is linear and
invertible, physical environment observables in the original basis can be
obtained by applying the inverse transformation on the chain basis used in
the numerical implementation.
In this section, we discuss how a tree tensor network structure provides
significant memory savings compared to the ``naive'' star network topology
discussed above. In the star network, the ``system'' ($\mathcal{S}$) is represented by a
tensor with $N+2$ dimensions (one physical index representing the coupled
exciton-photon state, as well as $N+1$ internal indices representing the
coupling to the environments, with maximum bond dimension $D$). This leads
to a severe memory bottleneck for large $N$ as the root tensor size scales
exponentially with $N$, $\mathcal{O}(D^{N+1})$.

In order to circumvent this exponential scaling while maintaining precision, we
decompose the system into a tree tensor network state~\cite{Szalay2015s,
	Shi2006s}, where each final branch is coupled only to a single chain (see
\autoref{fig:chain_mapping_tree}). This introduces additional auxiliary tensors
with no physical indexes, called ``entanglement renormalization
tensors''~\cite{Vidal2007s} which take in the complete chain states 
and pass on a reduced number (joint states) to the system.
In general, developing an efficient tree model
requires an explicit analysis of the entanglement topology of the state, in
essence analyzing possible regroupings and decompositions of bond legs over the
star-tensor-network. The condition for this compression to be effective is that 
there are correlations between the chains, i.e., the sum of the reduced state
entropy of each chain is greater than the joint entropy of the chains.
This idea has recently been implemented to allow the
simulation of multi-environment linear vibronic models constructed from ab
initio parametrizations of small molecules~\cite{Schroder2017Multis}. However, in
our specific case, permutation symmetry between the (identical) molecules holds.
An efficient tensor network is thus given by a structure with no privileged
distribution of phononic chains, i.e., the perfect binary tree in
\autoref{fig:chain_mapping_tree} with $\zeta$ levels for $2^\zeta$ molecules.
For simplicity, the environment $\mathcal{E}_\mathrm{r}$ is introduced as a
tensor chain connected directly to the root node, as the additional leg does not
increase memory storage critically.

Once the quantum wavefunction and the global Hamiltonian are represented in
tree-form, the time-dependent variational principle
algorithm~\cite{Haegeman2011s} can be implemented, generalizing the
single-chain algorithm~\cite{Haegeman2016s} by recursively optimizing each of
the child tensors of a given node in the tree-tensor network, and
environmental chains once the leaves of the tree are reached. More details
on this approach can be found in~\cite{Schroder2017Multis, Schroder2017Tensors}.

\section{Ensemble size effects}

In this section, we discuss the effect of changing the number of molecules
in the time evolution. To this effect, the system populations initialized in
the vibration-free upper polariton $\ket{+}$ for $N=4$ and $N=16$ molecules
are shown in \autoref{fig:N4_vs_N16}. In all cases, the increased number of
dark states when increasing $N$ leads to more efficient population transfer,
both when driven through off-resonant and multi-phonon processes (for
$\Omega_R=1~$eV, \autoref{fig:N4_vs_N16}a), and when a vibrational
transition is (close to) resonant with transitions between polaritons and
dark states ($\Omega_R=0.5~$eV and $\Omega_R=0.3$~eV,
\autoref{fig:N4_vs_N16}b-c). An analytical calculation of the rates in the
Markovian limit (the Bloch-Redfield-Wangness approach in the secular
approximation) predicts that, for a Rabi frequency below the vibrational
cutoff ($\Omega_R<\omega_c$), the global rate for a transition from the
upper polariton into the dark-state subspace $\mathcal{D}$ scales as
$(N-1)/N$, while the rate for the transitions $\rho_{++} \to \rho_{--}$ and
$\rho_{\mathcal{D}\mathcal{D}} \to \rho_{--}$ is suppressed as $\sim
1/N$~\cite{DelPino2015Quantums}. However, our results in \autoref{fig:N4_vs_N16}b,c
indicate that dark-state decay happens at comparable rates for $N=4$ and
$N=16$. Additionally, the time-dependent oscillation pattern is quite
similar for $N=4$ and $N=16$, with the threefold oscillation $\rho_{++}
\leftrightarrow \rho_{\mathcal{D}\mathcal{D}} \leftrightarrow \rho_{--}$
determined by the phononic reaction coordinate dynamics, but largely
independent of ensemble size. For $\Omega_R=0.1$~eV, shown in
\autoref{fig:N4_vs_N16}d, the ``universality'' in the early-time oscillation
frequency is preserved, but in contrast to the larger Rabi splittings,
the rate at which dark states decay into the lower polariton again decreases
with $N$. We interpret this as due to the breakdown of strong coupling,
which leads to the initial state $\ket{+}$ having contributions from
strongly vibrationally-dressed dark states which only decay inefficiently.

\section{Comparison with Holstein--Tavis--Cummings model}

\begin{figure}[tb]
	\centering
	\includegraphics[width=1\linewidth]{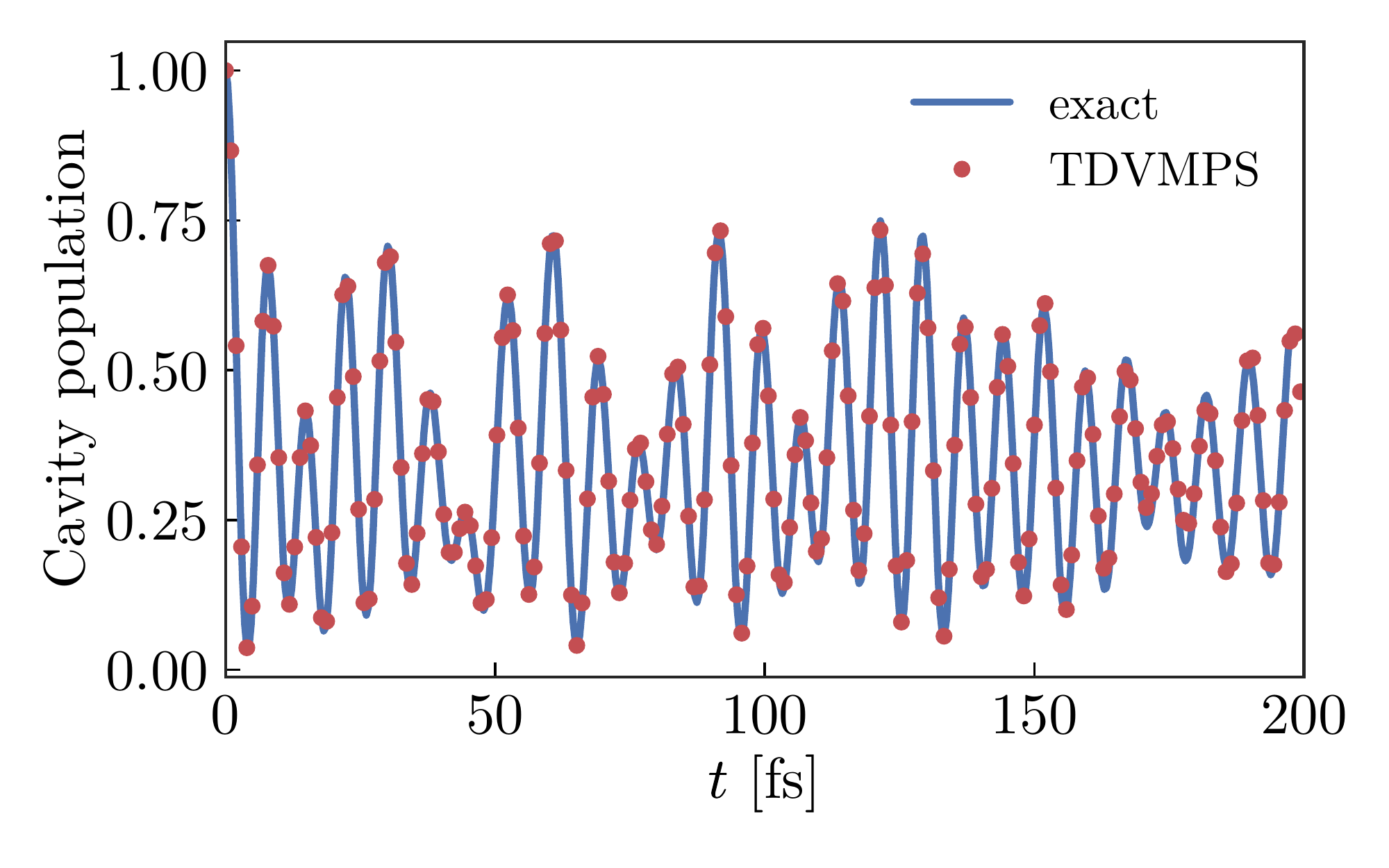}
	\caption{Comparison of the cavity population dynamics under the single-phonon mode description of $N=2$ molecules (HTC) between exact time propagation and a TDVMPS calculation. For this test, $\Delta=0.112$~eV and $\tilde{\omega}_1=0.154$~eV.}\label{fig:EXACT_vs_TDVMPS}
\end{figure}

\begin{figure}[b]\setlength{\hfuzz}{1.1\columnwidth}
	\begin{minipage}{\textwidth}
		\centering
		\includegraphics[width=1\linewidth]{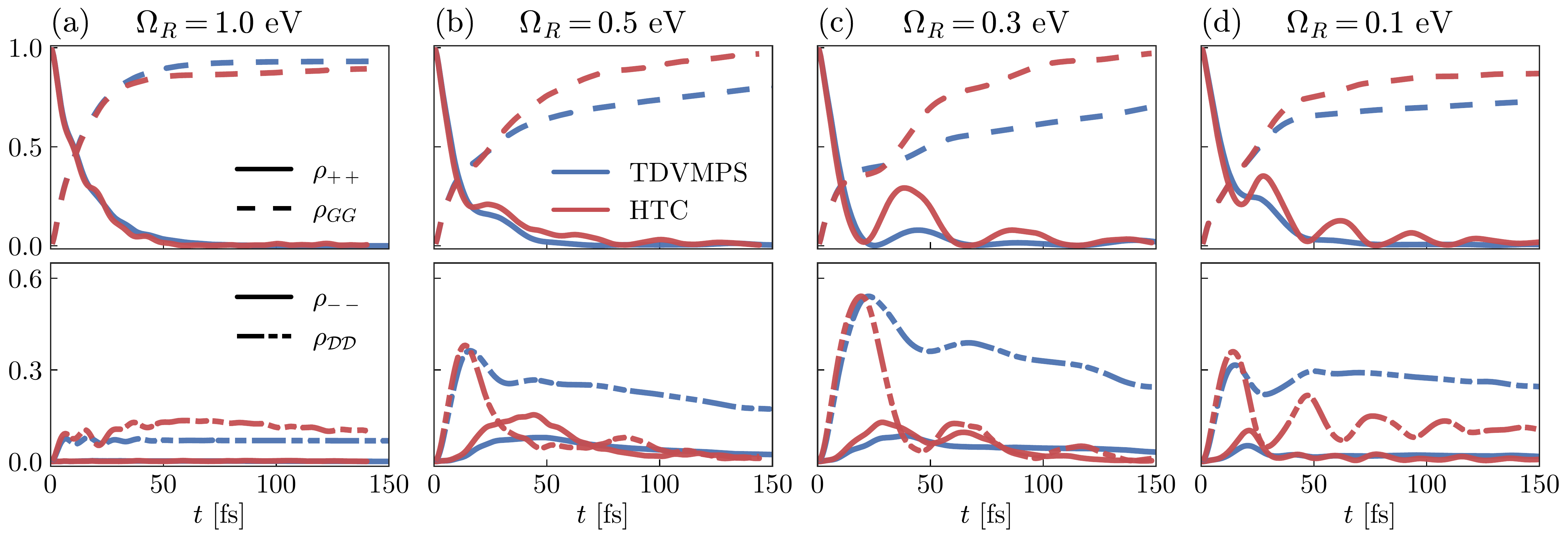}
		\caption{Comparison of population dynamics for $N=16$ between the full
			model (blue) and a single-phonon-mode description of the molecules
			(HTC, red). Parameters as in
			\autoref{fig:N4_vs_N16}.}\label{fig:N16_vs_HTC}
	\end{minipage}
\end{figure}


In this section we first check the reliability of the numerical method by comparing the time evolution of the loss-less Holstein-Tavis-Cummings (HTC) model, which only includes a single vibrational mode per molecule, via TDVMPS algorithm, with the result arising from an exact computation of the time propagation, retrieved via the open source library QuTiP~\cite{Johansson2013s}. The parameters of the HTC model are chosen to reproduce the reaction coordinate frequency of the Rh800 molecule
($\omega_\mathrm{HTC}=\tilde{\omega}_1=0.181$~eV) and total reorganization
energy $\lambda_\mathrm{HTC} = \sqrt{\omega_\mathrm{HTC}\Delta}$, with
$\Delta=\sum_k \lambda_k^2/\omega_k = 0.0356$~eV. This mapping has been found to
reproduce the most accurate lower phonon-polariton
state~\cite{DelPino2018Grounds}. 

As displayed by the reduced population $\rho_{11}$ in \autoref{fig:EXACT_vs_TDVMPS}, the TDVMPS time evolution almost exactly (to within the linewidth of the plot) reproduces the oscillatory features arising from the light-matter coupling and vibronic-induced effects in the exact calculation. This motivates the extension of the  approach to the exact multi-mode dynamics discussed in the main text, a regime where exact time propagation becomes unfeasible in practice. We address in the following the question of whether the full many-mode time dynamics of the system can be accounted for by means of the simplistic HTC model with parameters described above. As seen in \autoref{fig:N16_vs_HTC}a, while the HTC model reproduces the 
initial dynamics in the first few fs reasonably well (where the reaction
coordinate could be assumed to dominate the collective response), it
consistently overestimates the coherent oscillations observed for times
larger than about $10~$fs. In particular, it fails to correctly predict the
excitation trapping in the dark state subspace and in contrast leads to
enduring oscillations that are not dissipated into $\mathcal{D}$ but only
lost into photons. We have checked that choosing different
parameters (e.g., $\lambda_\mathrm{HTC}=\eta$) does not improve the agreement
significantly (not shown). 

\vspace{147mm}

\section{Convergence tests}\label{sec:conv}

Here we provide convergence checks of the tree-tensor network simulations, using the reduced population for the excited cavity ($\rho_{11}$) and vacuum states ($\rho_{GG}$) as benchmark observables. The relevant parameters in this analysis are \textit{i)} the maximum bond dimension of the tensor network $D$, \textit{ii)} the length of the chains for environments $\mathcal{E}_\mathrm{v},\mathcal{E}_\mathrm{r}$, denoted $L$ (and set equal to the number of photonic/phononic modes $L=M_v=M_r$), \textit{iii)} the time-step $\Delta t$ of time evolution, and finally \textit{iv)} the smallest singular value kept along the calculation, denoted as $\mathrm{sv}^{\mathrm{tol}}$. For reference, the values chosen in the main text are  $D=20$, $L=350$, $\Delta t=0.1$ eV$^{-1}$ (0.066 fs) and $\mathrm{sv}^{\mathrm{tol}}=10^{-4}$. In the following analysis, we will sweep each of these parameters separately, leaving the rest at these given values. In addition, it will be convenient to define the maximum relative error between the two solutions $\rho_{GG}^{\mathrm{sol}_1},\rho_{GG}^{\mathrm{sol}_2}$ where a single parameter is varied, $\epsilon=\max_t|\rho_{GG}^{\mathrm{sol}_1}(t)-\rho_{GG}^{\mathrm{sol}_2}(t)|/\rho_{GG}^{\mathrm{sol}_2}(t)$. 

\begin{figure}[tb]
	\centering
	\includegraphics[width=1\linewidth]{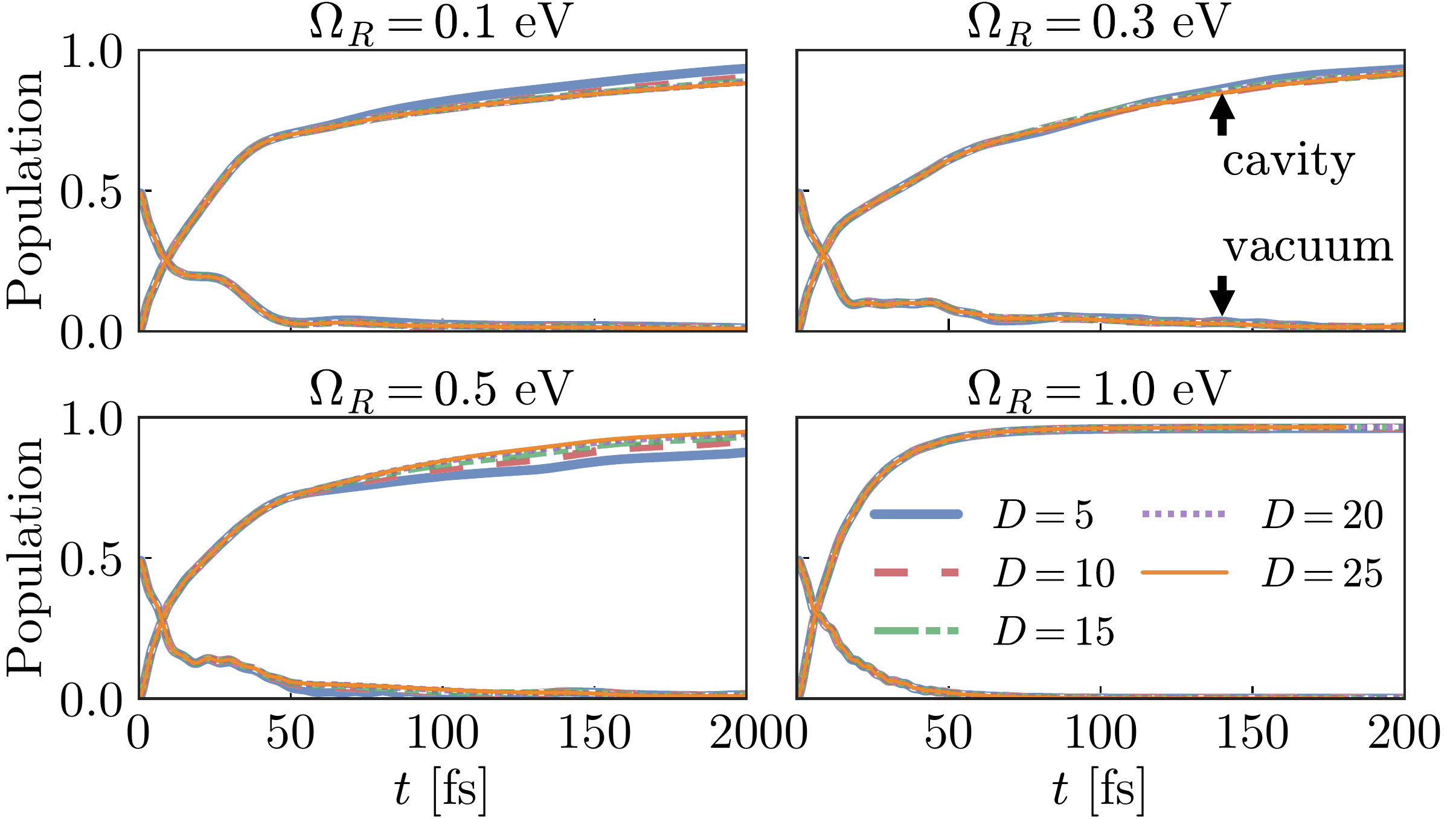}
	\caption{Comparison of population dynamics for $N=4$ between the full model with vaying bond dimension $D$. Other parameters as in \autoref{fig:N4_vs_N16}.}\label{fig:conv_D}
\end{figure}

In particular, $D$ corresponds to the number of ‘auxiliary’ states that encode the quantum correlations between neighboring degrees of freedom, thus setting a cutoff for the maximum entanglement entropy allowed in a given bond between two physical or entanglement renormalization nodes ($S_{\mathrm{max}}\sim\log D$~\cite{Schollwock2011s}). In \autoref{fig:conv_D}, we find acceptable results for $D>15$ for all Rabi frequencies $\Omega_R$. Convergence of populations is remarkably more demanding in terms of $D$ for the cases $\Omega_R=0.5$ and $0.1$ eV, where precisely Markovian (Bloch-Redfield-Wangsness) and non-Markovian time dynamics present stronger relative deviations (see main text). This observation establishes a direct link between the large amount of system-environment correlations and non-Markovianity in the time evolution. Moreover, analysis of the relative errors between the case ($D=20$), analyzed in the main text, and the best-converged case ($D=25$) shows a maximum deviation of $\epsilon<1\%$ during the first $200$ fs. 

Maximum dimensions $D^{\mathrm{OBB}}$ are also set for the bonds between chain tensors (for environments $\mathcal{E}_\mathrm{v}^{(i)},\mathcal{E}_r$) and the matrices mapping the optimal boson basis (OBB) into physical bosonic states~\cite{Guo2012s}. In practice, a value of $D^{\mathrm{OBB}}=50$ is sufficient to converge the dynamics in the main text, while total phonon populations in the output of the calculation (sum over chain occupations) stay typically below 1 for all cases analyzed in the main text. 

\begin{figure}[tb]
	\centering
	\includegraphics[width=1\linewidth]{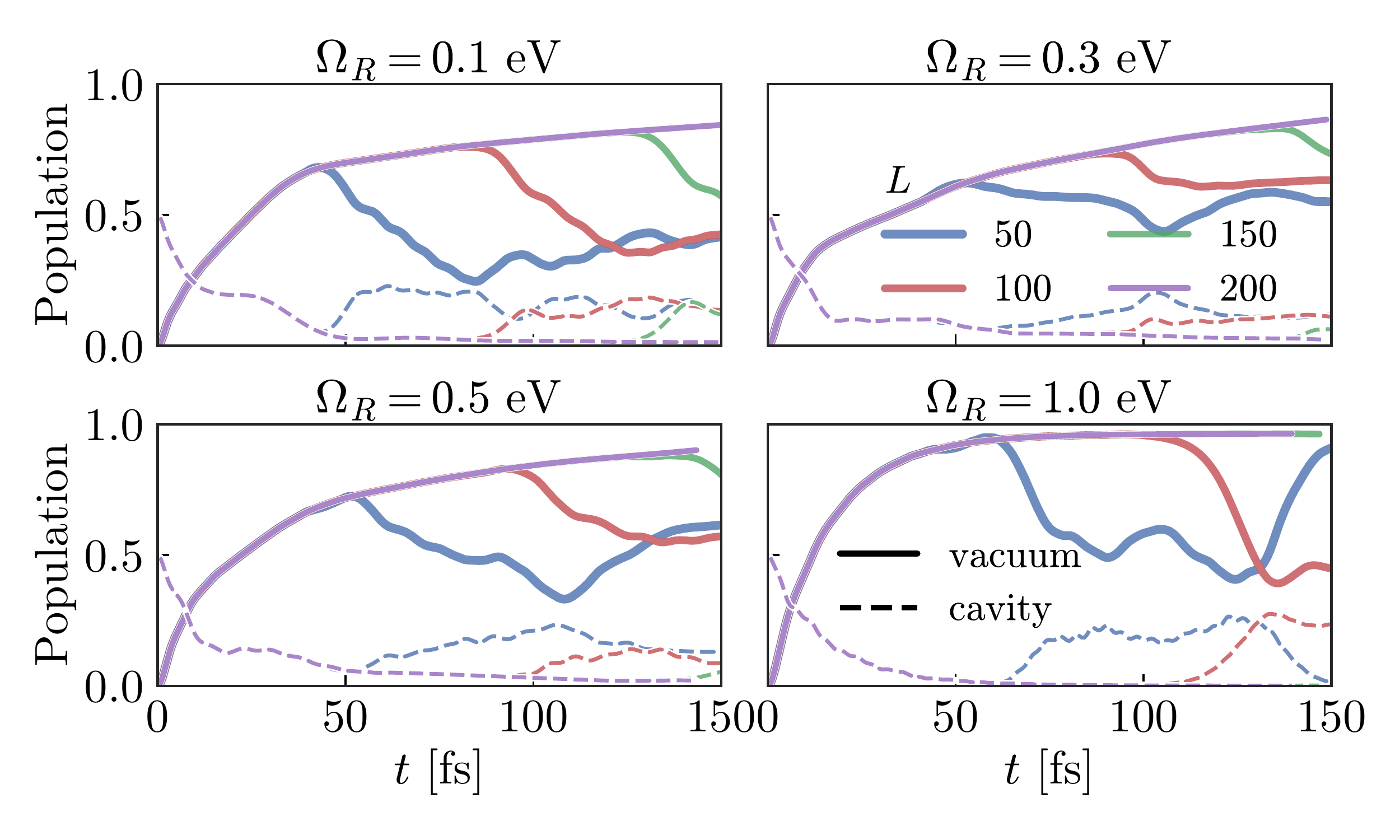}
	\caption{Comparison of population dynamics for $N=4$ for different chain lengths $L$ for environments $\mathcal{E}_\mathrm{v},\mathcal{E}_\mathrm{r}$. Other parameters as in \autoref{fig:N4_vs_N16}.}\label{fig:conv_L}
\end{figure}

Similar results of the time dynamics as a function of the chain length $L$ in \autoref{fig:conv_L} show artifacts in the density matrix $\hat{\rho}_\mathcal{S}$ at a time that can be estimated empirically to be $\sim L$ fs, with a weak dependence on the Rabi frequency. These are caused by the arrival of the (unphysically) reflected waves at the end boundaries of photonic and phononic finite chains, which is reminiscent of TDVMPS simulations of the spin-boson model~\cite{Schroder2016Simulatings}. In particular, inspection of the chain populations for different environments reveals that the limiting factor is the faster velocity group (steeper dispersion curve in chain wavevector space) of photonic wavepackets. In contrast with time evolution of system observables, environmental dynamics is profoundly sensitive to the finite boundaries (e.g. reflection of a photonic excitation implies an artificial breakdown of irreversible emission dynamics after the inverse chain mapping), demanding of the order of twice this chain length to calculate populations properly. In particular, to retrieve the emission spectrum dynamics up to $t>100$ fs in the main text, we employ a large value  $L=350$, preventing finite-size effects in the simulations.

\begin{figure}[tb]
	\centering
	\includegraphics[width=1\linewidth]{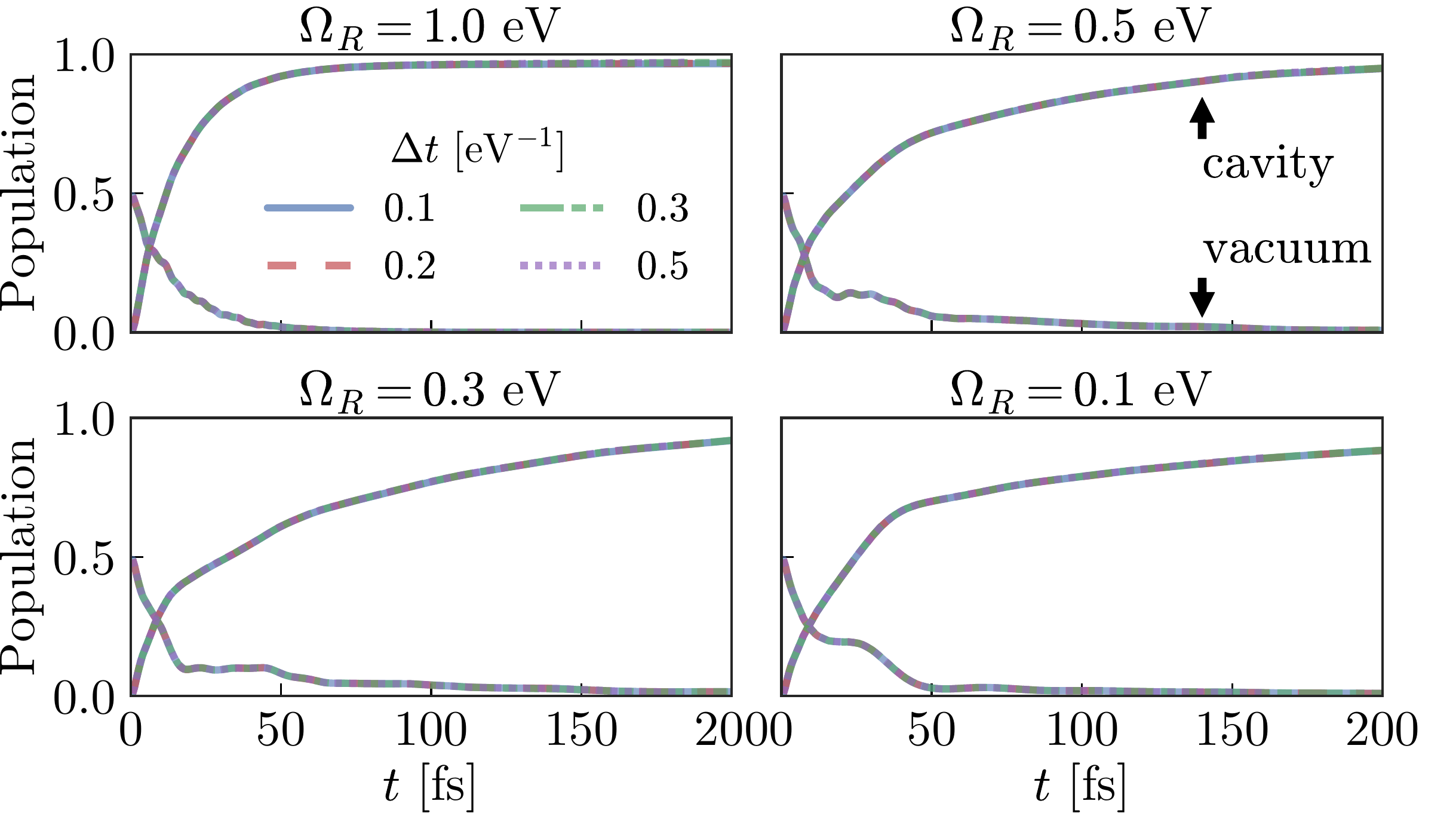}
	\caption{Comparison of population dynamics for $N=4$ for decreasing time step $\Delta t$. Other parameters as in \autoref{fig:conv_D}.}\label{fig:conv_dt}
\end{figure}

During the simulation, the many-body state is constructed after each
time-step $\Delta t$ after application of the full time evolution operator.
The error accrued in this propagation is discussed in detail in
Ref.~\cite{Haegeman2016s} and its references, which show that the error
arises only from the numerical method employed to integrate the TDVP
equations ($\mathcal{O}(\Delta t^3)$ for a left-right sweep along a single
chain)~\cite{Haegeman2011s,Schroder2016Simulatings}. In addition, it has
recently been pointed out that TDVMPS is accurate for local degrees of
freedom at very long times even with reduced $D$, as the projection
technique (leading to the TDVP approach) yields an effective Hamiltonian
that respects underlying conservation laws in the
system~\cite{Leviatan2017s}. In order to provide a quantification of the
error accrued during a global update of the tensor network by $\Delta t$,
convergence plots as a function of the time-step are shown in
\autoref{fig:conv_dt}. Here $\rho_{GG}$ quickly approaches the
asymptotically converged value for $\Delta t<0.4$ eV$^{-1}$ (0.33 fs) (see
\autoref{fig:conv_dt}), with relative errors that can be lowered below
$\epsilon=0.1\%$ for $\Delta t=0.1$ eV$^{-1}$ (value in main text), for
times shorter than 200 fs. 

\begin{figure}[htb]
	\centering
	\includegraphics[width=1\linewidth]{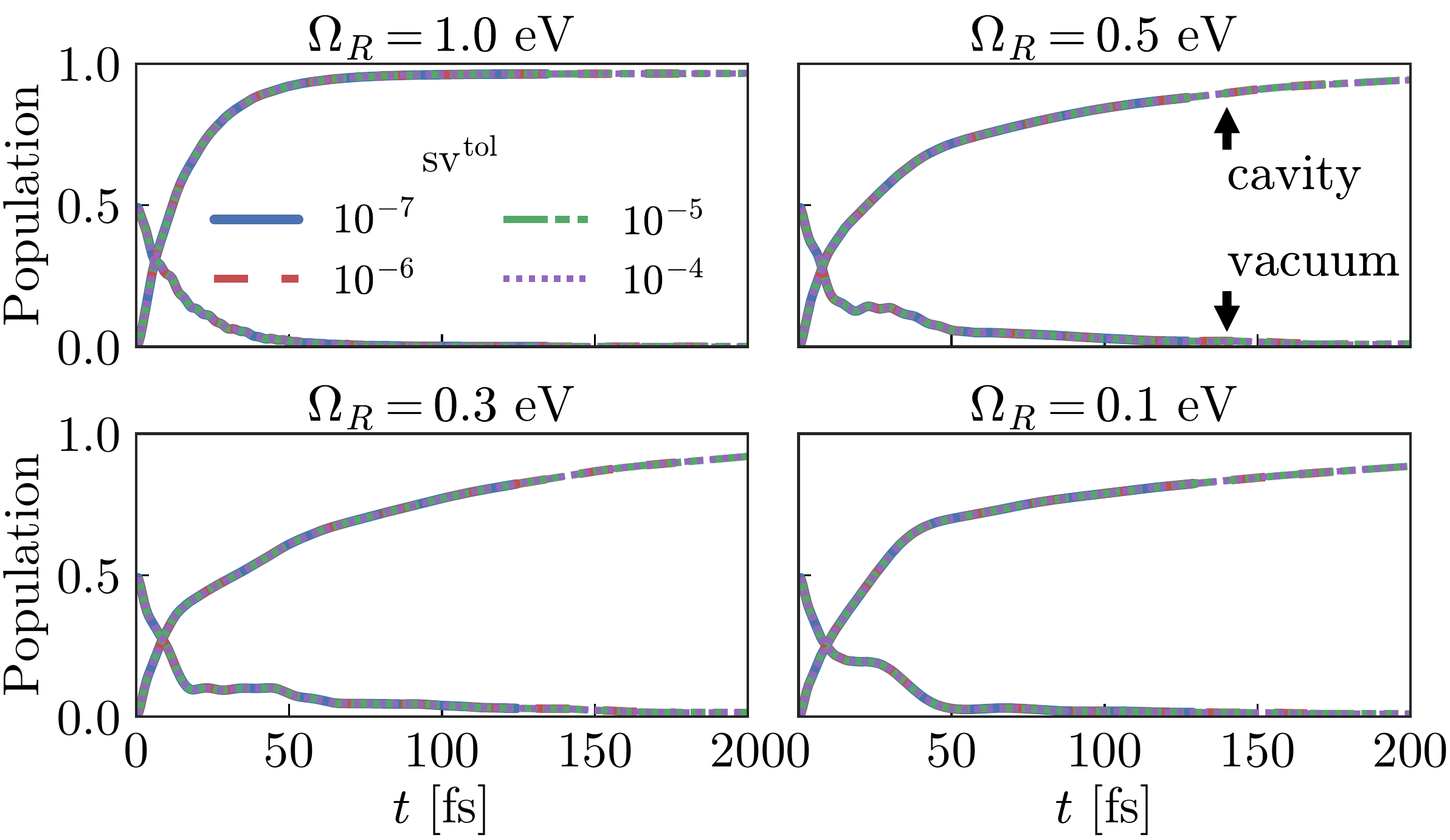}
	\caption{Comparison of population dynamics for $N=4$ for decreasing tolerance for singular value decomposition. Parameters as in \autoref{fig:N4_vs_N16}.}\label{fig:conv_SVD}
\end{figure}

In the algorithm implementation, bond dimensions are truncated or expanded
adaptively, according to the criteria that singular values $\{{\nu}_l\}$,
which measure the entanglement between either physical or entanglement
renormalization nodes (entanglement entropy
$S=-\sum_l|\nu_l|^2\log(|\nu_l|^2)\leq S_{\mathrm{max}}$), are truncated
below a value $\mathrm{sv}^{\mathrm{tol}}$, such that only the dominant
configurations required to reproduce the entangled many-body wave function
are kept during the calculation. The truncation scheme above is employed
similarly to the bonds between optimal boson basis matrices and the chain
tensors. \autoref{fig:conv_SVD} reveals converged populations for any value
below $\mathrm{sv}^{\mathrm{tol}}=10^{-4}$ (value for main text
calculations), and relative error analysis shows relative maximum deviations
of the state populations on the order of $\epsilon\sim1\%$ in comparison
with the most compute-intensive case, where
$\mathrm{sv}^{\mathrm{tol}}=10^{-7}$, for times shorter than 200 fs. 

\end{document}